1  **Assessing the consistency between short-term global temperature trends in**

2  **observations and climate model projections**

4  Patrick J. Michaels
5  George Mason University
6  Fairfax, Virginia
7  pmichaels@cato.org
8
9  Paul C. Knappenberger
10 New Hope Environmental Services, Inc.
11 Charlottesville, Virginia
12 chip@nhes.com
13
14 John R. Christy
15 Earth System Science Center, NSSTC
16 University of Alabama in Huntsville
17 Huntsville, Alabama
18 john.christy@nsstc.uah.edu
19
20 Chad S. Herman
21 Manalapan, New Jersey
22 chad.herman.us@gmail.com
23
24 Lucia M. Liljegren
25 Lisle, Illinois
26 lucia@rankexploits.com
27
28 James D. Annan
29 Research Institute for Global Change
30 Yokohama, Japan
31 jdannan@jamstec.go.jp

**Abstract**

Assessing the consistency between short-term global temperature trends in observations and climate model projections is a challenging problem. While climate models capture many processes governing short-term climate fluctuations, they are not expected to simulate the specific timing of these somewhat random phenomena—the occurrence of which may impact the realized trend. Therefore, to assess model performance, we develop distributions of projected temperature trends from a collection of climate models running the IPCC A1B emissions scenario. We evaluate where observed trends of length 5 to 15 years fall within the distribution of model trends of the same length. We find that current trends lie near the lower limits of the model distributions, with cumulative probability-of-occurrence values typically between 5% and 20%, and probabilities below 5% not uncommon. Our results indicate cause for concern regarding the consistency between climate model projections and observed climate behavior under conditions of increasing anthropogenic greenhouse-gas emissions.

**1. Background**

While global warming is often described as accelerating, in fact, the rate of increase in global average surface temperatures has slowed in recent years. However, the significance of this slowdown has not been well-established as most discussions about the issue lack sufficient grounding in the full distribution of the expectations to which the observations are being compared. Recent research has begun to focus on this issue, but has only done so in a limited scope. *Easterling and Wehner* [2009] determined the probability distribution for projected trends from a collection of climate models, but limited their analysis to trends of 10 years in length, while *Knight et al.* [2009] looked at the projected ranges for a variety of trend lengths, but from only one climate model.

Here we extend the results of previous analyses to determine the probability distribution of short-period trends in global temperature (in length from 5 to 15 years) as projected by a collection of climate models run under the Intergovernmental Panel on Climate Change (IPCC) A1B ("business-as-usual") emissions scenario. We then evaluate where the current values of the observed trends of similar length fall within the model distributions.

**2. Data and Methods**

*2.1 Climate Model Projections*

Monthly output from 20 climate models (51 model runs) incorporated in the IPCC *Fourth Assessment Report* [2007] run under the IPCC's A1B emissions scenario [*Nakićenović and Swart, 2000*] was obtained from Coupled Model Intercomparison Project 3 (CMIP3) [*Meehl et al., 2007*] database archived at the Program for Climate Model Diagnosis and Intercomparison (PCMDI) at the Lawrence Livermore National Laboratory. From these model projections, monthly global-average anomalies of surface and lower troposphere temperature were developed (see Auxiliary Material).

The model average temperature trend is very consistent for all trend lengths within the first two decades of the 21$^{st}$ century but begins to increase in the decades immediately thereafter. We therefore limit our analysis to the period January 2001 through December 2020 and consider this period to represent the expected behavior of the observed global average temperature during the first two decades of the 21$^{st}$ century.

Since the model runs contain internal (random) climate variability in addition to a response to the prescribed changes in radiative forcing, trends in model projections cannot be expected to match trends in observations over relatively short time spans—a few years to a decade or two. However, climate models do capture many characteristics of the primary processes driving short-term variability [*IPCC*, 2007, Chapter 8]. Therefore, the distribution of short-term temperature trends (of all lengths) from model projections should with high probability encompass the trends (of similar length) in the observed data if the model projections are accurately capturing climate behavior. While the observed trend falling within the model distribution of trends is not conclusive proof of the validity of climate model projections, it does serve as a necessary condition.

We develop the distributions of projected short-term temperature trends both for the surface and the lower troposphere. Through each individual model run, we calculate the moving linear trends through the first 20 years of monthly projections for time periods with lengths ranging from 5 years (60 months) to 15 years (180 months). For each model run, we develop the set of all available trends of each length. For example, for 5-year trends, we calculate the trend for the period January 2001-December 2005, February 2001-January 2006, March 2001-February 2006, successively stepping one month at a time thorough all 60-month periods and ending with January 2016-December 2020. The total number of trends determined from each model run declines with the increasing trend length, from 180 5-year trends, to 60 15-year trends. For each trend length, we then combine the set of trends calculated from each of the 51 model runs—weighted to produce an equal contribution from each climate model (regardless of the number of available runs)—into a single distribution representing a sample of the overall population of potential realities contained in the collection of climate models [*Annan and Hargreaves*, 2010]. Weighting each model run equally does not materially affect our results. The distribution of 5-yr trends contains contributions from 9,180 (180 x 51) elements, a number which declines to 3,060 for 15-yr trends (60 x 51). However, all individual elements are not independent of each other as the moving trends within a single model run are to some degree correlated.

*2.2 Observed Temperature Record*

We use observed records of global average surface temperature anomalies compiled monthly by the Climate Research Unit of the University of East Anglia and the Hadley Centre (HadCRU) [*Brohan et al.*, 2006], by the Goddard Institute for Space Studies (GISS) [*Hansen et al.*, 2006] and by the National Climatic Data Center (NCDC) [*Smith et al.*, 2008]. Additionally we use observed records of global average lower troposphere temperatures measured by Microwave Sounder Units (MSU) aboard satellites as complied by the University of Alabama-Huntsville (UAH) [*Christy et al.*, 2003] and by Remote Sensing Systems (RSS) [*Mears and Wentz*, 2009].

From the observed global temperature anomalies in each dataset, we calculate the linear trends using simple least squares regression of lengths 5 years (60 months) to 15 years (180 months) ending with the most recent data available (December 2009) (see Auxiliary Table 1 for the observed trend values).

Observed trends of length greater than 9 years include data from a period of time prior to the IPCC AR4 climate model projections (which generally begin in January 2001). However, the rate of increase of radiative forcing from anthropogenic emissions changes very little between the mid-1990s and the first few decades of the 21$^{st}$ century under the A1B emissions scenario [*IPCC*, 2007] so a comparison between observed behavior over the past 15 years and the model expected behavior during the period 2001-2020 is appropriate. We do not extend our analysis into trends of length greater than 15 years as the observed trend begins to be influenced by the 1991 eruption of Mt. Pinatubo—a type of natural forcing not included in the A1B emissions scenario.

## 3. Results and Discussion

There are several options to assess the cumulative probability of a particular trend value within the model distributions of projected trends. For instance, the cumulative probability of a 10-yr trend in global average surface temperatures with a value less than

146  or equal to zero can be determined directly from the elements of the distribution of model
147  projected 10-yr trends by using ranked percentiles (which yields a cumulative probability
148  of 6.3%), by using Student's t-distribution conservatively with 31 degrees of freedom
149  representing the weighted combination of the 51 model runs (which yields a cumulative
150  probability of 8.4%), or by fitting a normal distribution (which yields a cumulative
151  probability of 7.9%). The results of these three solutions are very similar across all trend
152  lengths, indicating that the determination of the cumulative probability is not overly
153  sensitive to the choice of method. As such, subsequently we will only report the results
154  using the assumption of normality.

155

156  These results in the previous example can be compared with other assessments of model
157  trend probabilities. *Easterling and Wehner* [2009] used a similar statistical methodology,
158  but used model projections from the SRES A2 scenario to determine the probability of a
159  10-yr trend less than or equal to zero. They reported a probability of "about 10%" for
160  such an occurrence during the first half of the 21$^{st}$ century. This value is slightly greater
161  than the value from our methodology, mostly likely, because the A2 scenario examined
162  by *Easterling and Wehner* [2009] includes less forcing during the first half of the 21$^{st}$
163  century than does the A1B scenario we used. *Knight et al.* [2009] examined variability
164  within the trends produced by the HadCM3 climate model when run under a variety of
165  emissions scenarios and model settings. *Knight et al.* [2009] found that a 10-yr trend falls
166  just inside the 90% range of trends produced by the HadCM3 model—a value apparently
167  similar to ours.

168

In Figure 1 we present a general depiction of the model probability distributions for trends of length 5 to 15 years for surface temperatures. As the length of the trend increases, the probably range tightens. This general solution can be used to assess the model-based probability of any and all short-term trends within the first 20 years of the 21$^{st}$ century. For example, the probability of a trend in global average temperatures that is less than or equal to zero becomes 5% or less at a length of about 11 years (132 months). The probability distributions for the projected trends in the lower troposphere are very similar (see Auxiliary Figure 1). The average model projected trend in the lower troposphere is about 20% larger than the surface (0.025°C/yr vs. 0.020°C/yr) and the spread about the mean is slightly larger as well.

The spread of the distributions of model projected trends is governed both by statistical uncertainty about the best-fit linear trend that results from random variability that is independent from month-to-month, as well as by the influence of random (over the longer-term) low-frequency variability that is correlated over times scales of months to decades and which may alter the value of the short-term trends for an extended time period. Our working hypothesis is that these random processes operate to influence model trends to the same degree as they do observed trends. Therefore, we assume that the model trend distributions represent the spread of potential realities (including these uncertainties), of which the single realization of the observed trend is a member.

One notable exception to this assumption concerns the true observational errors, such as those arising from incomplete spatial coverage, station number changes, and non-climatological influences on the temperature measurements. These errors do not occur in the model projections for which the temperature is precisely known. Estimates of the size of observational errors are available for each observed dataset and we incorporate them into Monte Carlo simulations to ascertain their influence on variability of trends ranging from 5 to 15 years in length. We add this variability to the variability in the model trend distributions (see Auxiliary Material). This results in a slight broadening of the distributions.

From these adjusted distributions, derived separately for the surface and the lower troposphere, we determine the cumulative probably of occurrence of the value of the observed trend (ending in December 2009) ranging in length from 5 to 15 years in each of the five observed datasets—three compilations of surface temperatures and two compilations of lower tropospheric temperatures (Figure 2).

The cumulative probabilities of the observed trend values typically are less than 20% (with the exception of GISS dataset). In all datasets the cumulative occurrence probability of the current 8-yr trend is about 10% or less, and in all datasets except the GISS dataset, there is less than a 10% probability of current values for trends of 7, 8, 9, 12, and 13 years in length. The values for these same trend lengths from some datasets fall beneath the 5% cumulative probability indicating an expectation of occurrence of less than 1 in 20 (a typical measure of statistical significance). In general, the cumulative probabilities of the observed trends are lower for the lower troposphere than for the surface.

## 4. Conclusions

For most observational datasets of global average temperature, the trends from length 5 to 15 years lie along the lower tails of the probability distributions from the collection of climate model projections under the SRES A1B emissions scenario. Typically the probability of occurrence of the observed trend values lies between 5% and 20%, depending on the dataset and the trend length. In the HadCRU, RSS, and UAH observed datasets, the current value of trends of length 8, 12, and 13 years is expected from the models to occur with a probability of less than 1 in 20. Taken together, our results raise concern about the consistency between the observed evolution of global temperatures in recent years and the climate model projections of that evolution.

Possible reasons for why current trends are unusual when set among model projections include unknown errors in the observational temperature record, differences in the true vs. A1B-defined anthropogenic forcing changes, insufficiencies of the climate models to accurately replicate the characteristics of natural variability, inaccuracies in climate model transient climate evolution, and the overestimation by climate models of the actual climate sensitivity. These are in addition to the possibility that current trends represent simply a rare but not impossible situation that is generally captured by the climate models.

As global emissions of carbon dioxide—the primary anthropogenic climate forcing agent—have been increasing during recent years at a rate similar to that specified in the A1B scenario [*Nakićenović and Swart*, 2000; *EIA*, 2008], it is unlikely that the difference between observed and projected trends arises from a significant underestimate of the changes in climate forcing prescribed by the A1B scenario. Similarly, while there are clearly differences among the observed trend values derived from the various observational datasets, all trends through the observed data fall in the lower tails of model projections, so it is unlikely that errors in the observations (which may include a warming bias in surface observations in recent years, [e.g., *McKitrick and Michaels*, 2007; *Klotzbach et al.*, 2009] are the primary cause of the observed/projected differences. This leads to the conclusion that a large part of the differences between the observed trends and model-projected trends lies with the internal workings of the models. This conclusion is supported by results which indicate that natural variations in ocean/atmospheric circulation patterns are in part responsible for the recent slowdown in the rate of global temperature rise [*Keenlyside et al.*, 2008; *Swanson and Tsonis*, 2009] and that inadequately-modeled decadal-scale variations in stratospheric water vapor have a significant influence on global temperature trends, including contributing to a reduced trend in recent years [*Solomon et al.*, 2010]. Further, some results indicate that the model determinations of climate sensitivity may be too large [e.g., *Wyant et al.*, 2006; *Spencer and Braswell*, 2008]. It can also be noted that the discrepancy between observed trends and projected trends is greater for satellite than surface observations.

259  Our results stand in contrast to results such as *Rahmstorf et al.* [2007] which concluded

260  that observed trends through global average temperatures are increasing at a rate near the

261  upper end of the IPCC projected range. The primary reasons for the contrasting

262  conclusions are that our analysis is based upon updated climate model runs, more recent

263  observed data, and a more comprehensive analysis of model projections.

264

265  **6. References**

266

267  Annan, J. D., and J. C. Hargreaves (2010), Reliability of the CMIP3 ensemble. *Geophys.*

268  *Res. Lett.*, *37*, L02703, doi:10.1029/2009GL041994.

269

270  Brohan, P., J. J. Kennedy, I. Harris, S. F. B. Tett and P.D. Jones (2006), Uncertainty

271  estimates in regional and global observed temperature changes: a new dataset from 1850,

272  *J. Geophys. Res., 111*, D12106, doi:10.1029/2005JD006548.

273

274  Christy J. R., R. W. Spencer, W. B. Norris, W. D. Braswell, and D. E. Parker (2003),

275  Error estimates of version 5.0 of MSU–AMSU bulk atmospheric temperatures. *J. Atmos.*

276  *Oceanic Technol., 20,* 613–629.

277

278  Easterling, D. R., and M. F. Wehner (2009), Is the climate warming or cooling? *Geophys.*

279  *Res. Lett.*, *36*, L08706, doi:10.1029/2009GL037810.

280

281    Energy Information Administration (EIA) (2008), *International Energy Annual, 2006*.

282    U.S. Department of Energy, Washington, D.C.,

283    http://www.eia.doe.gov/pub/international/iealf/tableh1co2.xls

284

285    Hansen, J., M. Sato, R. Ruedy, K. Lo, D. W. Lea and M. Medina-Elizade (2006), Global

286    temperature change, *Proc. Natl. Acad. Sci.*, *103*, 14288-14293,

287    doi:10.1073/pnas.0606291103.

288

289    Intergovernmental Panel on Climate Change (IPCC) (2007), *Climate Change 2007: The*

290    *Physical Science Basis. Contribution of Working Group I to the Fourth Assessment*

291    *Report of the Intergovernmental Panel on Climate Change*, edited by S. Solomon et al.,

292    996 pp., Cambridge Univ. Press, Cambridge, U. K.

293

294    Keenlyside, N. S., M. Latif, J. Jungclaus, L. Kornblueh and E. Roeckner (2008),

295    Advancing decadal-scale climate prediction in the North Atlantic sector. *Nature*, *453*, 84-

296    88, doi:10.1038/nature06921.

297

298    Klotzbach, P. J., R. A. Pielke Sr., R. A. Pielke Jr., J. R. Christy, and R. T. McNider

299    (2009), An alternative explanation for differential temperature trends at the surface and in

300    the Lower Troposphere, *J. Geophys. Res.*, *114*, D21102, doi:10.1029/2009JD011841.

301

302    Knight, J., Kennedy, J. J., Folland, C., Harris, G., Jones, G. S., Palmer, M., Parker, D.,

303    Scaife, A., & Stott, P. (2009), Do global temperature trends over the last decade falsify climate predictions? In: Peterson, T. C., & Baringer, M.O. (eds), "State of the Climate in 2008" Special Supplement to the *Bull. Am. Meteorol. Soc.*, 90-91.

McKitrick, R. R., and P. J. Michaels (2007), Quantifying the influence of anthropogenic surface processes inhomogeneities on gridded global climate data. *J. Geophys. Res., 112*, D24S09, doi:10.1029/2007JD008465.

Mears, C. A., and F. J. Wentz (2009), Construction of the RSS V3.2 lower tropospheric temperature dataset from the MSU and AMSU microwave sounders. *J. Atmos. Ocean. Tech., 26*, 1493-1509.

Meehl, G. A., et al. (2007), The WCRP CMIP3 multi-model dataset: A new era in climate change research, *Bull. Am. Meteorol. Soc., 88*, 1383-1394.

Nakićenović, N., and R. Swart (2000), *Intergovernmental Panel on Climate Change Special Report on Emissions Scenarios*, Cambridge Univ. Press, Cambridge, U. K.

Rahmsdorf, S., A. Cazenave, J. A. Church, J. E. Hansen, R. F. Keeling, D. E. Parker, and R. C. Somerville (2007), Recent climate observations compared to projections. *Science, 316*, 709.

Smith, T.M., R. W. Reynolds, T. C. Peterson, and J. Lawrimore (2008), Improvements to NOAA's historical merged land-ocean surface temperature analysis (1880-2006), *J. Clim.*, *21*, 2283-2296.

Solomon, S., K. H. Rosenlof, R. W. Portmann, J. S. Daniel, S. M. Davis, T. J. Sanford, and G-K Plattner (2010), Contributions of stratospheric water vapor to decadal changes in the rate of global warming, *Science*, *327*, 1219-1223.

Spencer, R. W., and W. D. Braswell (2008), Potential biases in feedback diagnosis from observations data: a simple model demonstration, *J. Clim., 21*, 5624-5628.

Swanson, K. L. and A. A. Tsonis (2009), Has the climate recently shifted? *Geophys. Res. Lett., 36,* L06711, doi:10.1029/2008GL037022.

Wyant, M. C., M., Khairoutdinov, and C. S. Bretherton (2006), Climate sensitivity and cloud response of a GCM with a superparameterization. *Geophys. Res. Lett.*, *33*, L06714.

**Figure Captions**

Figure 1. Cumulative probability distribution of trend values for trends ranging in length from 5 to 15 years derived from 20 models under SRES A1B for the period January 2001 through December 2020 for global average surface temperatures. The 95% confidence range is shaded in grey and a zero trend is indicated by the horizontal black line.

Figure 2. Cumulative probabilities of the current observed values of the trends ranging in length from 5 to 15 years (each ending in December 2009) through average global surface temperature anomalies and lower troposphere temperature anomalies as complied within five observed temperature datasets.

Figure 1.

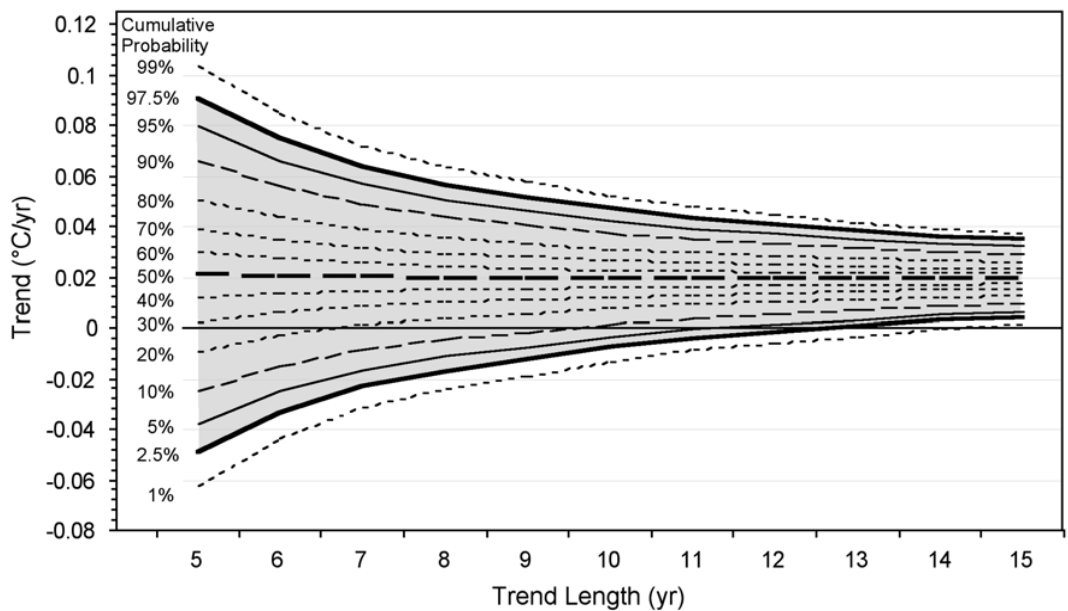

Figure 1. Cumulative probability distribution of trend values for trends ranging in length from 5 to 15 years derived from 20 models under SRES A1B for the period January 2001 through December 2020 for global average surface temperatures. The 95% confidence range is shaded in grey and a zero trend is indicated by the horizontal black line.

Figure 2.

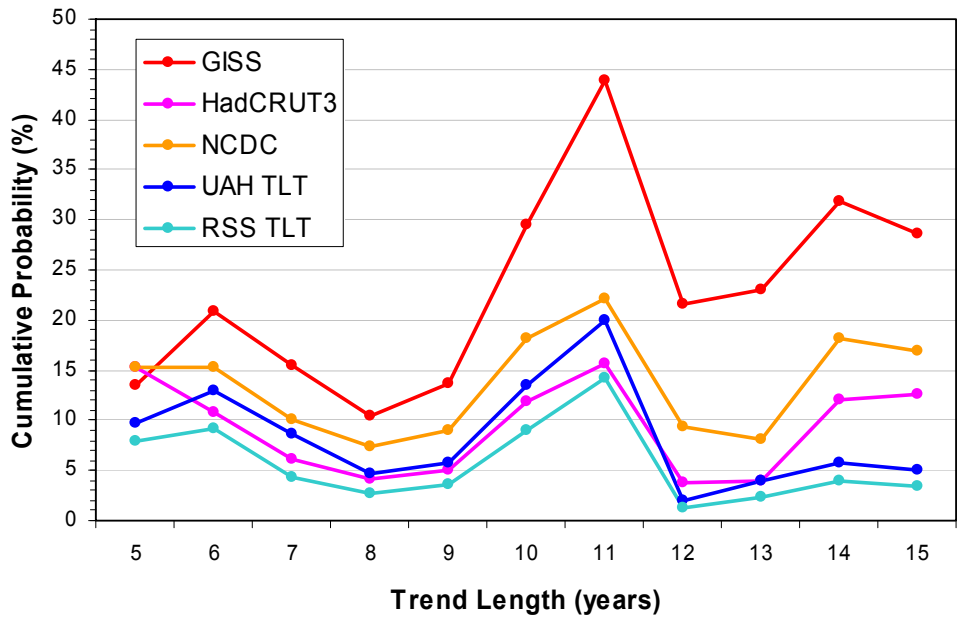

Figure 2. Cumulative probabilities of the current observed values of the trends ranging in length from 5 to 15 years (each ending in December 2009) through average global surface temperature anomalies and lower troposphere temperature anomalies as complied within five observed temperature datasets.

Auxiliary material for paper …

**Assessing the consistency between short-term global temperature trends in observations and climate model projections**

Patrick J. Michaels
George Mason University
4400 University Drive, Fairfax, Virginia 22030
pmichaels@cato.org

Paul C. Knappenberger
New Hope Environmental Services, Inc.
536 Pantops Center, #402
Charlottesville, Virginia 22911
chip@nhes.com

John R. Christy
Earth System Science Center, NSSTC
University of Alabama in Huntsville
Huntsville, Virginia 35805
john.christy@nsstc.uah.edu

Chad S. Herman
Marlboro, New Jersey
chad.herman.us@gmail.com

Lucia. M. Liljegren
Lisle, Illinois
lucia@rankexploits.com

James D. Annan
Research Institute for Global Change
Yokohama, Japan
jdannan@jamstec.go.jp

39  *Auxiliary Methods*

41  *1. Climate Model Selection*

42  Gridded monthly projections of surface and atmospheric air temperatures from 51 individual climate model runs representing 20 separate climate models were downloaded from the PCMDI CMIP3 model archive database. Model selection consisted of only those listed in IPCC AR4 Table 10.4. and runs whose 20C3M runs continue forward with SRES A1B. Of these models, two had to be eliminated. There was no atmospheric temperature data available for MIUB ECHO-G so lower troposphere temperatures could not be simulated. CNRM CM 3 was eliminated because the netCDF files did not represent the atmospheric temperature on a consistent set of pressure levels.

51  The models and the numbers of runs we used in our analysis are included in Table 1 (details of these climate models can be found at the PCMDI archive, http://www-pcmdi.llnl.gov/ipcc/model_documentation/ipcc_model_documentation.php).

Table 1. Model names and number of available runs.

| Model Name | Number of Runs |
|---|---|
| BCCR BCM 2.0 | 1 |
| CCCMA CGCM 3.1 T47 | 5 |
| CCCMA CGCM 3.1 T63 | 1 |
| CSIRO MK 3.0 | 1 |
| ECHAM5/MPI-OM | 4 |
| GFDL CM 2.0 | 1 |
| GFDL CM 2.1 | 1 |
| GISS AOM | 2 |
| GISS EH | 3 |
| GISS ER | 5 |
| IAP FGOALS 1.0g | 3 |
| INM CM 3.0 | 1 |
| IPSL CM 4 | 1 |
| MIROC 3.2 HIRES | 1 |
| MIROC 3.2 MEDRES | 3 |
| MRI CGCM 2.3.2a | 5 |
| NCAR CCSM 3.0 | 7 |
| NCAR PCM 1 | 4 |
| UKMO HAD CM 3 | 1 |
| UKMO HADGEM 1 | 1 |

*2. Creation of monthly surface air temperature anomalies*

For each model run, the projected monthly gridded surface air temperature values were spatially averaged to produce global average temperatures for each month. The global average monthly temperatures were then converted to global average monthly temperature anomalies by subtracting the climatology for each model run over the period January 2001 through December 2020.

*3. Creation of monthly synthetic MSU lower troposphere temperatures anomalies*

Microwave Sounder Units (MSU) carried aboard a series of NASA satellites monitor bulk average temperatures in the atmosphere. A temperature for the lower troposphere can be generated from the MSU observations by a weighted combination of several MSU frequency channels. To properly compare the observed MSU lower troposphere temperatures with climate model projections, model-generated atmosphere temperature data must be used to develop an equivalent synthetic MSU lower troposphere temperature product. To this end, we employed the procedure described by Santer et al. [1999] as implemented by Santer and Doutriaux [2005] as part of the PCMDI Climate Data Analysis Tools package to produce gridded monthly, synthetic MSU lower troposphere

76  temperatures from each model run. The gridded temperature values were spatially
77  averaged to produce global average temperatures for each month. The global average
78  monthly temperatures were then converted to global average monthly temperature
79  anomalies by subtracting the climatology for each model run over the period January
80  2001 through December 2020.

*References*

84  Santer, B., and C. Doutriaux, 2005. Subroutine MSUWEIGHTS3.f, Part of the Climate
85  Data Analysis Tools available from the Lawrence Livermore National Laboratory's
86  (LLNL) Program for Climate Model Diagnosis and Intercomparison (PCMDI),
87  http://www2pcmdi.llnl.gov/svn/repository/cdat/tags/CDAT4.3/contrib/MSU/Src/msuwei
88  ght.f and archived at http://www.webcitation.org/5owusvk91.

92  Santer, B. D., J. J. Hnilo, T. M. L. Wigley, J. S. Boyle, C. Doutriaux, M. Fiorino, D. E.
93  Parker, and K. E. Taylor (1999), Uncertainties in observationally based estimates of
94  temperature change in the free atmosphere, J. Geophys. Res., 104(D6), 6305–6333.

*4. Accounting for "observational error"*

97  Observations of the average global temperature contain uncertainties that model projected
98  temperatures do not. These "observational errors" include such things as incomplete
99  spatial coverage, changing number of stations within gridcells, changing observational practices, etc. The magnitude of these errors has been quantified in the literature describing each of the observed datasets that we used in our study. For the UAH MSU lower troposphere temperatures, the standard errors for the monthly anomalies are given in Christy et al [2003]. Brohan et al. [2008] describes the monthly components of observational error that are contained in the HadCRUT3 surface temperatures. For the GISS [Hansen et al., 2006] and NCDC [Smith et al., 2008] surface temperatures, however, only the standard error of the annual anomalies are presented. The information quantifying the errors of monthly global anomalies in the RSS MSU lower troposphere temperatures was obtained through personal communications [Mears, 2010].

Since we are using monthly anomalies, we require estimates of the errors for monthly anomalies. The Hadley Center website (http://hadobs.metoffice.com/hadcrut3/diagnostics/global/nh+sh/) provides global temperature anomalies as well as error ranges for the HadCRUT3 dataset, for both monthly and annual anomalies. If the errors were independent from one another at the monthly timescale, the annual error would be equal to the monthly error divided by the square root of 12 (or by a factor of 3.46). However, comparing the listed monthly and annual error ranges, we find that the annual error in the HadCRUT3 is only reduced by a factor of 1.73 from the monthly errors (or the square root of 3), indicating that the errors are not independent (Table 2). The major source of observational error in the global temperature anomalies complied in the HadCRUT3 data is a result of incomplete spatial coverage, with a minor contribution from bias [Brohan et al., 2008]. The error resulting from bias has a lower-frequency variability than does the error from incomplete spatial coverage [Brohan et al., 2008].

Since we have only estimates of the error of the annual global anomalies available from the GISS [Hansen et al., 2006] and the NCDC [Smith et al., 2008] datasets, we will use the HadCRUT monthly-to-annual scaling factor to guide our estimation of the error about the monthly anomalies reported in the NCDC and GISS datasets. Smith et al. [2008] finds that in the NCDC dataset, the contribution from bias is greater than the contribution from incomplete spatial coverage, as they use interpolation to increase the spatial coverage of the observations. Since bias error is more temporally correlated than error resulting from incomplete spatial coverage, we reduce the scaling factor that we determined from the HadCRUT3 data from 1.73 down to 1.50 for the NCDC dataset to account for the likely reduced degrees of freedom in the NCDC error compared with HadCRUT3 errors. As the characteristics of the spatial coverage of the GISS data are similar to that of the NCDC data, we apply the 1.50 scaling factor to the GISS data as well. The reported annual error, along with our estimated monthly error for these datasets is listed in Table 2.

Table 2. The standard error of "observational errors" of the annual and monthly global temperature anomalies from the 5 observed datasets used in our analysis.

| Dataset   | Annual Error (°C) | Monthly Error (°C) |
| --------- | ----------------- | ------------------ |
| HadCRUT3* | 0.045             | 0.078              |
| NCDC      | 0.03              | 0.045              |
| GISS      | 0.025             | 0.0375             |
| UAH MSU   | 0.075             | 0.10               |
| RSS MSU   | 0.043             | 0.047              |

150  *This is an average for the HadCRUT3 errors (over 1979-2009) as the actual errors are
151  computed monthly and differ from month to month

153  To assess the influence of these observational errors on the trends of length 5 to 15 years

154  ending in December 2009 in each dataset, we used a Monte Carlo simulation, drawing

155  each monthly data element randomly from a normal distribution with a mean equal to the

156  observed monthly global temperature anomaly, and a standard deviation equal to the

157  monthly error listed in Table 1 (for the HadCRUT3 data, we used the error explicitly as

158  reported on the Hadley Center web site,

159  http://hadobs.metoffice.com/hadcrut3/diagnostics/global/nh+sh/, which differs from

160  month to month). We performed 10,000 replications for each series (from 5 to 15 years in

161  length ending in December 2009) for each dataset, determining the linear least-squares

162  trend through each. From the distributions of 10,000 values for each trend length for each

163  dataset, we determined the standard deviation representing the trend variability due to

164  observational errors assuming independence from month to month.  The effect of

165  correlations in the monthly errors was not assessed. The likelihood that the correlations in

166  monthly errors vary considerably in time and across datasets makes it difficult to

167  speculate whether the trend variability would be higher or lower than we determined

168  assuming error independence on the specific 5 to 15 years trends in this study.

170  We incorporated the influence of observational into our distributions of model trends by

171  adding the standard deviation from observational error determined for each trend length

172  and each observational dataset as described above, in quadrature to the standard deviation

173  of the model trend distributions of the same length, $\sigma_{x+y} = \text{sqrt}(\sigma_x^2+\sigma_y^2)$, where $\sigma_{x+y}$ is the

174 combined standard deviation, $\sigma_x^2$ is the standard deviation of the model trend
175 distributions, and $\sigma_y^2$ is the standard deviation of observational error. Each distribution of
176 modeled trends was broadened by the inclusion of the effect of observational errors.
177
178

179  Auxiliary Table 1. Value of the linear least-squares trend for lengths ranging from 5 years
180  (60 months) to 15 years (180 months) through global average temperature anomalies
181  ending in December 2009, from the five observed datasets used in this study.
182

| Trend Length (yrs) | Trend (°C/yr) | | | | |
|---|---|---|---|---|---|
| | GISS | HADCRUT3 | NCDC | UAH MSU v5.2 | RSS MSU v3.2 |
| 5 | -0.01835 | -0.01608 | -0.01564 | -0.03560 | -0.04033 |
| 6 | -0.00178 | -0.01409 | -0.00773 | -0.01573 | -0.02256 |
| 7 | -0.00205 | -0.01445 | -0.00782 | -0.01451 | -0.02412 |
| 8 | -0.00365 | -0.01325 | -0.00735 | -0.01615 | -0.02147 |
| 9 | 0.00169 | -0.00761 | -0.00242 | -0.00797 | -0.01232 |
| 10 | 0.01214 | 0.00292 | 0.00689 | 0.00542 | 0.00154 |
| 11 | 0.01783 | 0.00733 | 0.01037 | 0.01203 | 0.00882 |
| 12 | 0.01099 | -0.00001 | 0.00511 | -0.00283 | -0.00522 |
| 13 | 0.01224 | 0.00226 | 0.00590 | 0.00395 | 0.00149 |
| 14 | 0.01556 | 0.00957 | 0.01189 | 0.00873 | 0.00715 |
| 15 | 0.01526 | 0.01061 | 0.01217 | 0.00937 | 0.00785 |

183

184

185

186

187   Auxiliary Figure 1.

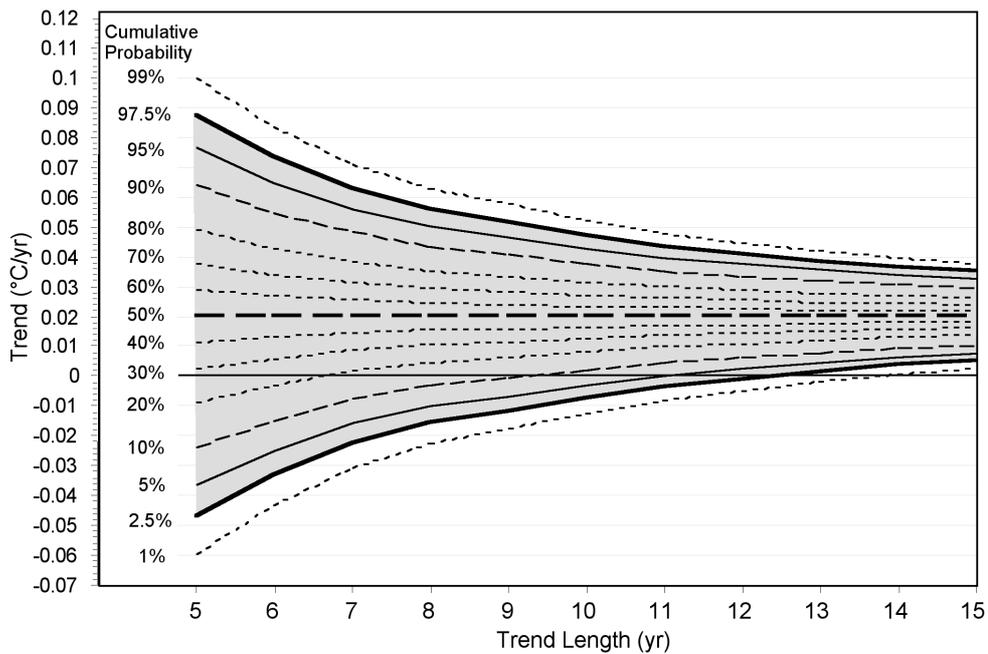

Auxiliary Figure 1. Cumulative probability distribution of trend values for trends ranging in length from 5 to 15 years derived from 20 models under SRES A1B for the period January 2001 through December 2020 for global average MSU lower troposphere temperatures. The 95% confidence range is shaded in grey and a zero trend is indicated by the horizontal black line.